\documentclass[pra,showpacs,twocolumn,superscriptaddress]{revtex4}

\usepackage{amsfonts}
\usepackage{amstext}
\usepackage{amssymb,amsmath,graphicx,enumerate,color}
\usepackage{hyperref}

\newcommand{\ket}[1]{\left| #1 \right\rangle}
\newcommand{\bra}[1]{\left\langle #1 \right|}

\newcommand{\op}[1]{{\mathbf{#1}}}

\newcommand{\doubint}{\int \!\!\!\!\int}

\newcommand{\eg}{{\it{e.g.}}: }
\newcommand{\ie}{{\it{i.e.}}: }

\newcommand{\erf}{\mathrm{erf}}

\begin{document}

\title{Increasing identical particle entanglement by fuzzy measurements }
\author{D. Cavalcanti}\email{dcs@fisica.ufmg.br}
\affiliation{Departamento de F\'{\i}sica - CP 702 - Universidade Federal de Minas
Gerais - 30123-970 - Belo Horizonte - MG - Brazil}
\author{M. Fran\c ca Santos}\email{msantos@fisica.ufmg.br}
\affiliation{Departamento de F\'{\i}sica - CP 702 - Universidade Federal de Minas
Gerais - 30123-970 - Belo Horizonte - MG - Brazil}
\author{M.O. \surname{Terra Cunha}}\email{tcunha@mat.ufmg.br}
\affiliation{Departamento de Matem\'atica - CP 702 - Universidade
Federal de Minas Gerais - 30123-970 - Belo Horizonte - MG -
Brazil}
\author{C. Lunkes}\email{christian.lunkes@ic.ac.uk}
\affiliation{QOLS, Blackett Laboratory, Imperial College London, London SW7 2BZ,
England}
\author{V. Vedral}\email{v.vedral@leeds.ac.uk}
\affiliation{Institut f\"ur Experimentalphysik, Universit\"at Wien,
Boltzmanngasse 5, A-1090 Vienna, Austria} \affiliation{The School of
Physics and Astronomy, University of Leeds, Leeds LS2 9JT, England}

\begin{abstract}
We investigate the effects of fuzzy measurements on spin entanglement for
identical particles, both fermions and bosons. We first consider an ideal
measurement apparatus and define operators that detect the symmetry of the
spatial and spin part of the density matrix  as a function of particle
distance.  Then, moving on to realistic devices that can only detect the
position of the particle to within a certain spread, it was surprisingly found
that the entanglement between particles increases with the broadening of
detection.
\end{abstract}

\maketitle

Entanglement is one of the most studied subjects in contemporary
physics. This quantum mechanical property is the core of a whole
new area in physics, mathematics and computation, known as quantum
information theory. It has also been perceived as a useful
resource for different technological applications~\cite{nielsen}.
However much has still to be done in order to completely
understand these non-classical correlations. One of the important
aspects that still deserves further investigation concerns the
actual observation of entanglement due to the indistinguishable
character of the quantum world.

Usually, when talking about entanglement, one tends to ignore the role of the
measurement apparatus, always considering ideal situations. However, there is no such
a thing as an ideal detector, and the detector bandwidth affects the measurement of
entanglement. Furthermore, in the particular case of identical particles, it is still
not clear how the symmetry of detection in external degrees of freedom affects the
entanglement of the internal ones.

In this paper we analyze these questions in two different systems:
a gas of non-interacting fermions at zero temperature and a
photonic interferometer. In particular, we analyze how the
measurement of external degrees of freedom can affect the
entanglement in internal degrees of freedom in a system of
identical particles. First, we show how the symmetry of detection
affects the entanglement of two fermions detected in a
non-interacting fermion gas at zero temperature. Then, we study
how imperfect detections affect the observation of entanglement
both for the fermionic gas and also for an interferometer of
photons.

Recently Vedral showed that the entanglement between two particles in a
non-interacting Fermi gas at zero temperature decreases with increasing
distance from each other~\cite{vedral}. Vedral also shows that there is a limit
below which any two fermions extracted from the gas are certainly entangled.
In particular, if both fermions are extracted at the same position, then the
Pauli exclusion principle forces their spin to be maximally entangled in a
typical antisymmetric Bell state.

In order to calculate the upper limit for guaranteed entanglement, the author
used the Peres-Horodecki criterion~\cite{peres}, which in his paper,
corresponds to $f^2>\frac{1}{2}$, where $f(r-r')=3j_1(k_F |r-r'|)/(k_F|r-r'|)$,
$|r-r'|$ is the distance between the fermions, $p_f=\hbar k_F $ is the Fermi
momentum and $j_1$ is a spherical Bessel function. In order to obtain this
result, the author defines (apart from normalization) the spin density matrix
of these two particles from the second order correlation function (see the Appendix for a justification)
\begin{equation}
\rho^{(2)}_{ss',tt'}=\bra{\Phi_0} \Psi^\dagger_{t'}(r')\Psi^\dagger_t(r)\Psi_{s'}(r')
\Psi_s(r)\ket{\Phi_0}, \label{dm1}
\end{equation}
which corresponds to measuring these two fermions at positions $r$
and $r'$ (the subscriptions $s$,$s'$, $t$, and $t'$ refer to the
spin of the particles). The initial state $| \phi_0 \rangle$ is the
ground state of the Fermi system
\begin{equation}
\ket{\phi_0}=\Pi_{s,p}b_{s}^\dag (p)\ket{0},
\end{equation}
and the detection operator at position $r$ and spin $s$ is given by,
\begin{equation}
\Psi_{s}^\dag(r)=\int_{0}^{p_f} \frac{d^3 p}{(2\pi)^3}e^{-ipr}b_{s}^\dag(p).
\label{psi}
\end{equation}
A typical interpretation of this result is related to the spatial overlap of
the wavefunctions of each fermion.

Here, we present a clearer and more complete explanation to this behavior
through the analysis of the symmetry of the position detection. To do so let us
define new detection operators:
\begin{subequations}
\begin{eqnarray}
\Pi^+_{ss'}(r,r')&\equiv &\frac{\left(\Psi_{s}(r)
\Psi_{s'}(r')+\Psi_{s'}(r)\Psi_s(r')\right)}{\sqrt{2}}, \label{pi+}\\
\Pi^-_{ss'}(r,r')&\equiv &\frac{\left(\Psi_{s}(r)
\Psi_{s'}(r')-\Psi_{s'}(r)\Psi_s(r')\right)}{\sqrt{2}}.
\end{eqnarray}
\end{subequations}
The operator $\Pi^+_{ss'}(\Pi^-_{ss'})$, detects the antisymmetric (symmetric)
spatial part and the symmetric (antisymmetric) spin part of  fermion wavefunction.

In terms of these new operators, Eq.(\ref{dm1})
becomes:\begin{widetext}
\begin{equation}
\rho_{ss',tt'}^{(2)}=\frac{1}{2}\langle \Phi_0|
[\Pi^{+\dagger}_{tt'}(r,r')+\Pi^{-\dagger}_{tt'}(r,r')]
[\Pi^{+}_{ss'}(r,r')+\Pi^{-}_{ss'}(r,r')]|\Phi_0\rangle
\label{dm2}
\end{equation}\end{widetext}
Note that written in this way, the two-fermion density matrix is the sum of
four different terms, two of which contain only symmetric and antisymmetric
spin detectors. The other two are the crossing terms, which vanish due to the
exclusion principle (spin and position of two fermions cannot be both symmetric
or antisymmetric). The remaining terms are:
\begin{subequations}
\begin{eqnarray}
\rho_{sym}&=&\frac{1}{2}\bra{\Phi_0} \Pi^{+\dagger}_{tt'}(r,r')\Pi^{+}_{ss'}(r,r')
\ket{\Phi_0}, \label{dm2a}\\
\rho_{asym}&=&\frac{1}{2}\bra{\Phi_0} \Pi^{-\dagger}_{tt'}(r,r')\Pi^{-}_{ss'}(r,r')
\ket{\Phi_0}. \label{dm2d}
\end{eqnarray}
\end{subequations}
Here $\rho_{sym}$ takes into consideration only the symmetric spin function (therefore, it
is related to the detection of the antisymmetric part of the spatial wavefunction)
while $\rho_{asym}$ contains only the antisymmetric spin function (therefore related
to the detection of the symmetric part of the spatial wavefunction). The density
matrix can be rewritten as:
\begin{eqnarray}
\rho^{(2)}&=& \rho_{asym}+\rho_{sym} \nonumber
\\&=& \doubint dp dp' \frac{1}{2}\left\{
(1+e^{i(p-p')(r-r')})\left (\begin{array}{cccc}
   &  &  &  \\
   & 1 & -1 &  \\
   & -1 & 1 &  \\
   &  &  &
\end{array} \right)\right. \nonumber \\ &+&\left.
(1-e^{i(p-p')(r-r')})\left (\begin{array}{cccc}
  2 &  &  &  \\
   & 1 & 1 &  \\
   & 1 & 1 &  \\
   &  &  & 2
\end{array} \right) \right\}.
\label{dmf2}
\end{eqnarray}
First, note that, as expected, for $r=r'$, the antisymmetric spatial function
goes to zero, and the spin wavefunction has to be antisymmetric (first term in
Eq.(\ref{dmf2})). For $r-r' \neq 0$ both parts contribute. Note, however, that
the symmetric spin density matrix can be viewed as an equal weight mixture of
the three triplet components, and has no entanglement at all (by the
Peres-Horodecki criterion). The spin state in Eq.(\ref{dmf2}) represents a
convex combination of singlet state and the equal mixture of triplet states. It
will have entanglement iff the fraction of singlet is sufficiently larger than
the fraction of triplets, in a similar context to the calculation of
\emph{relative robustness}~\cite{robust}. In the limit $r-r' \gg 1/k_F$, as the
integrals on Eq.(\ref{dmf2}) are performed over momenta below the (momentum
equivalent of) Fermi surface the momentum dependent terms oscillate too fast,
and average to zero. Therefore, the spin density matrix becomes just the
identity ($\delta_{ts}\delta_{t's'}$). This behavior can be seen as a smooth
transition from a quantum statistics (Fermi-Dirac) to a classical one
(Maxwell-Boltzmann).

Another way of analyzing Eq.(\ref{dmf2}) is given in
Ref.\cite{ohkim}, where the authors describe $\rho^{(2)}$ as a
Werner state. Here, it becomes clear how the symmetry of the
spatial detection is influencing the behavior of the spins, and
also why there is no entanglement from $|r-r'| \sim 1/k_F$ on.

An interesting question arises when considering non-punctual
(non-ideal) position detections, i.e. more realistic apparatus that
detect the position of those fermions with some uncertain. Instead
of Eq.(\ref{psi}), the detection operator should be written as the
general field operator:
\begin{equation}
\Psi_s (r)= \doubint e^{ipr''}D(r-r'') b_s(p)dr'' dp.
\label{genpsi1}
\end{equation}
The perfect position measurement situation is the particular case of
Eq.(\ref{genpsi1}) corresponding to $D(r-r'')=\delta(r-r'')$. However, if
$D(r-r'')$ has some position uncertainty, like, for example, if it is described
by a gaussian with spread $\sigma$,
\begin{equation}
D(r-r'')=\frac{1}{\sqrt{2\pi}\sigma}
e^{\frac{-|r-r''|^2}{2\sigma^2}},\label{quale}
\end{equation}
then, the field operators become
\begin{equation}
\Psi_s(r)= \frac{1}{\sqrt{2\pi}\sigma}\doubint e^{ipr''}
e^{\frac{-|r-r''|^2}{2\sigma^2}}b_s(p)dr''dp.
\end{equation}
These operators, when substituted back in Eq.(\ref{dm1}), give:
\begin{subequations}
\begin{eqnarray}
\rho_{ss',tt'}&=&\delta_{st}\delta_{s't'}f(d,\sigma)+
\delta_{st'}\delta_{s't}g(d,\sigma),
\end{eqnarray}
with $d=|r-r'|$,
\begin{equation}
f(d,\sigma)=\left( \frac{\erf (\sigma p_f)}{\sigma}\right)^2,
\end{equation}
\begin{eqnarray}
g(d,\sigma)&=&\frac{1}{\sigma^2}e^{\frac{-d^2}{2\sigma^2}}\{\erf(\sigma
p_f-\frac{id}{2\sigma})-\erf(-\frac{id}{2\sigma})\}\nonumber
\\&\times&\{\erf(\sigma
p_f+\frac{id}{2\sigma})-\erf(\frac{id}{2\sigma})\},
\end{eqnarray}
and $\erf(x)$ is the ``error function'', defined via:
\begin{equation}
erf(x)\equiv\frac{2}{\sqrt{\pi}}\int_{0}^{x} exp(-t^2)dt.
\end{equation}
\label{dmsub}
\end{subequations}

\begin{figure}[h]
  \includegraphics[width=8cm]{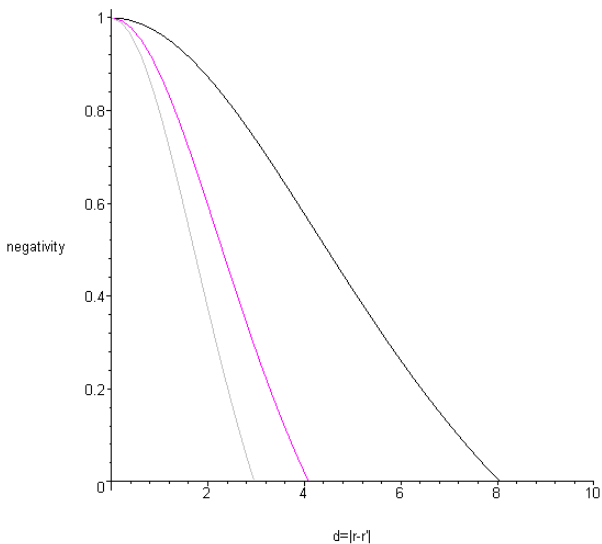}\\
  \caption{(Color online) $\mathcal{N}(\rho)$ vs. $d$ for $p_f =1$ and $\sigma=1$ (grey),
  $\sigma=2$ (magenta) and $\sigma=4$ (black).}\label{NEG}
\end{figure}

In order to make clearer the behavior of entanglement in relation to
changes in $d$ and $\sigma$ we can compute the negativity
($\mathcal{N}$)~\cite{negativity} of the state $\rho^{(2)}
(d,\sigma,p_f)$. This entanglement quantifier was chosen because it
is an \emph{entanglement monotone}~\cite{monot} (\ie satisfies all
the required features of a good entanglement quantifier), and also it is easy to
calculate. $\mathcal{N}(\rho)$ is equal to twice the absolute value
of the negative eigenvalue of the partial transpose of $\rho$,
$\rho^{T_A}$. $\mathcal{N}(\rho)$ measures by how much
$\rho^{T_A}$ fails to be positive semi-definite\footnote{It is
important to note that $\mathcal{N}(\rho)$ quantifies only
non-positive-partial-transpose states, NPPT-states. However it is
not a problem here because every 2-qubit entangled state is a
NPPT-state \cite{horod}.} \cite{negativity}. In our case
$\mathcal{N}(\rho)$ reads as
\begin{equation}
\mathcal{N}(\rho)=2\mu( \frac{f-2g}{4f-2g})
\end{equation}
where $\begin{array}{c}
 \mu=\left\{\begin{array}{rl}
 -x\ ,& x<0\\
 0\ , & x\geq0\
 \end{array}\right.
 \end{array}$.
This function is plotted in Fig. \ref{NEG} for some values
of $\sigma$.

Note that, for imperfect position detection, the entanglement decreases as the
detectors become apart from each other, but increases if the spread in the
detection becomes larger. The fact that inaccuracy in the detection increases
entanglement seems surprising. However it has to be noted that as our knowledge
in position gets worst, our knowledge in momentum gets better. In the limit of
infinite spread, both detectors
become perfect momentum detectors (centered at $p=0$, see Eq.\eqref{quale}),
 which means again that their spin wavefunction should be totally
antisymmetrized, hence they are found in the antisymmetric Bell state. It is important to stress that Eq.\eqref{genpsi1} describes a coherent combination of localized field operators instead of a statistical average of them. That is the reason for the infinite spread limit be a momentum-localized detector instead of just a vague ``there is a particle somewhere''.

A similar situation can be found in a bosonic system calculating the
entanglement in polarization between photons placed in an interferometer using
polarization beam splitters (PBS). In this analysis the polarization and
frequency of the photons play the role of the spin and momentum of the
fermionic case. This kind of interferometry has been extensively studied in
practical experiments \cite{hongoumandel} and thus, in principle, the results
presented here could be tested experimentally.

\begin{figure}[h]
  \includegraphics[width=8cm]{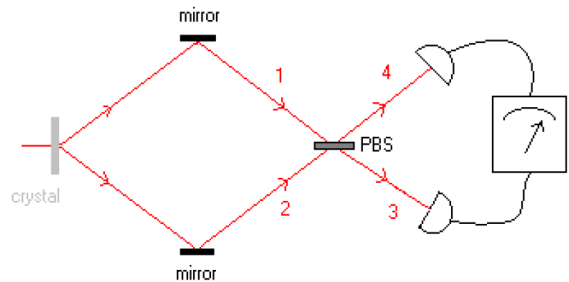}\\
  \caption{(Color online) Hong-Ou-Mandel type of interferometer.}\label{fig:interferometro}
\end{figure}

First we write the modes after the PBS if the modes before it are
$a_{1, \epsilon}$ and $a_{2, \epsilon}$ (see
Fig.\ref{fig:interferometro}):
\begin{subequations}
\begin{eqnarray}
E_{3,\epsilon}(\omega) &=& T_{\epsilon}a_{1, \epsilon}(\omega)+ i{R_{\epsilon}}a_{2,
\epsilon}(\omega)e^{i\omega \tau},\\
E_{4,\epsilon}(\omega) &=& iR_{\epsilon}a_{1, \epsilon}(\omega)+ T_{\epsilon}^{*}a_{2,
\epsilon}(\omega)e^{i\omega \tau},
\end{eqnarray}
\end{subequations}
with $\vert T_\epsilon \vert^{2}$ and $\vert R_\epsilon \vert^{2}$ being the
transmissivity and reflectivity of the $\epsilon$ polarization in the PBS, $\tau=\tau_1 -\tau_2$,
where $\tau_1$ and $\tau_2$ are the propagation times from the
crystal to the detectors via path 1 and path 2 respectively.

The two particle density matrix is again defined through the second-order correlation
function:
\begin{eqnarray}
 \rho_{\epsilon {\epsilon}', s
s'}&=&\doubint d\omega_1 d\omega_2 D(\omega_1 , \omega_2)
\\&& \times \bra{\phi_0}E_{3,s}^{\dag}(\omega_1)
E_{4,s'}^{\dag}(\omega_2)E_{4,\epsilon'}(\omega_2) E_{3,\epsilon}(\omega_1) \ket{\phi_0}
\nonumber.
\end{eqnarray}
The state
\begin{equation}
\vert \phi_0 \rangle=\psi^{\dag} \ket{0}
\end{equation}
is prepared by the two-photon creation operator
\begin{equation}
\psi^{\dag}=\sum_{k,k'} \doubint a_{1, k}^{\dag}(\nu_1) a_{2,
k'}^{\dag}(\nu_2) g_{k,k'}(\nu_1,\nu_2) d\nu_1 d\nu_2,
\end{equation}
with frequency distribution $ g_{k,k'}(\nu_1,\nu_2)$ which is, in the more
general case, polarization dependent. Note that this calculation is more
general than a typical Hong-Ou-Mandel interferometer, since one is free to
choose initial states by $g_{k,k'}(\nu _1,\nu _2)$. In a way similar to
Eq.\eqref{genpsi1}, detection is represented by a function $D(\omega _1, \omega
_2)$.

Using the commutation relation for bosons and evaluating the integrals in $d\nu_1,
d\nu_2$ and the summations in $k, k'$ we find:
\begin{widetext}
\begin{eqnarray}
\rho_{\epsilon {\epsilon}', s' s} &=&
 \doubint \{T_{s'} T_{s}^* T_{\epsilon'}^* T_\epsilon
g_{ss'}^* (\omega_1 ,\omega_2) g_{\epsilon \epsilon'} (\omega_1 ,\omega_2) +
R_{s'} R_{s}^* R_{\epsilon'}^* R_\epsilon g_{s's}^* (\omega_2 ,\omega_1)
g_{\epsilon' \epsilon} (\omega_2 ,\omega_1)  \\&-& T_{s'} T_{s}^* R_{\epsilon'}^* R_\epsilon g_{ss'}^*
(\omega_1 ,\omega_2)g_{\epsilon' \epsilon} (\omega_2,\omega_1)e^{i(\omega_1 - \omega_2) \tau} -
T_{\epsilon'}^* T_\epsilon R_{s'} R_{s}^* g_{\epsilon \epsilon'} (\omega_1,\omega_2)
g_{s's}^* (\omega_2 ,\omega_1)e^{-i(\omega_1 - \omega_2)\tau}\}  D(\omega_1 , \omega_2)
d\omega_1 d\omega_2 .\nonumber
\end{eqnarray}

The PBS is an optical element which allows transmission in only
one polarization (say vertical), and reflection in the other one
(horizontal). The coefficients of transmissivity and reflectivity
can be written as $R_h = T_v =1$ and $T_h = R_v =0$. With this,
and assuming a symmetric state in polarization ($g_{ss'}
=g_{s's}$), the unique non-vanishing coefficients are:
\begin{subequations}
\begin{eqnarray}
\rho_{\epsilon \epsilon,\epsilon \epsilon}&=&\doubint |g_{\epsilon \epsilon}(\omega_1 ,\omega_2)|^2
 D(\omega_1 , \omega_2)d\omega_1 d\omega_2, \ \ \epsilon = h,v, \\
\rho_{hh,vv}&=& - \doubint g_{vv}^* (\omega_1 ,\omega_2)g_{hh}(\omega_2
, \omega_1)e^{i(\omega_1 - \omega_2)} D(\omega_1 , \omega_2)d\omega_1 d\omega_2 = \rho ^*_{vv,hh}.
\end{eqnarray}
\end{subequations}
\end{widetext}

In order to carry on the calculations, let us assume particular
distributions for the functions $g(\omega_1 ,\omega_2)$ and $D(\omega_1
, \omega_2)$. Since we are more interested in the detection properties,
let us first assume that $g(\omega_1 ,\omega_2)$ is constant.
Then, we assume independent gaussian detectors, which means
$D(\omega_1, \omega_2)=D(\omega_1)D(\omega_2)$ and
\begin{equation}
D(\omega)=\frac{1}{\sqrt{2\pi}\sigma}e^{-\frac{\omega^2}{2\sigma^2}}.
\end{equation}
Now, we can calculate the polarization density matrix, obtaining, up to normalization,
\begin{equation}
\rho_{hh,hh}=\rho_{vv,vv}=\doubint D(\omega_1)D(\omega_2)d\omega_1 d\omega_2=1,
\end{equation}
and
\begin{eqnarray}
\rho_{hh,vv}&=&\rho^{*}_{vv,hh}=-\frac{1}{2\pi\sigma^2}\int
\exp(-\frac{\omega_{1}^2}{2\sigma^2}+i\omega_1 \tau)d\omega_1 \nonumber \\
&&\times\int \exp(-\frac{\omega_{2}^2}{2\sigma^2}-i\omega_2 \tau)
 d\omega_2.
\end{eqnarray}
Evaluating these Fourier transforms of Gaussian functions we obtain
\begin{equation}
\rho_{hh,vv}=\rho_{vv,hh}=-\exp(-\sigma^2\tau^2).
\end{equation}
The (normalized) polarization density matrix then is given by
\begin{equation}
\rho=\frac{1}{2}\left(
\begin{array}{cccc}
  1 & 0 & 0 & -f\\
  0 & 0 & 0 & 0 \\
  0 & 0 & 0 & 0 \\
  -f & 0 & 0 & 1  \\
\end{array}
\right),
\end{equation}
where $f=\exp(-\sigma^2\tau^2)$. The partial transposition of $\rho$ has always
one negative eigenvalue equal to $-\frac{f}{2}$. So the Negativity of $\rho$ is
$\mathcal{N}(\rho)=f=\exp(-\sigma^2\tau^2)$. This result shows that as the
detector becomes broader in frequencies or the time delay between the two arms
of the interferometer becomes larger, the entanglement between detected photon
pairs decreases. One should note that time delay gives rise to which path
information, while broadening in frequencies opposes to such labeling.

Again we can see the state $\rho$ as a combination of Bell states.
Interestingly, in the interferometric process described here the
separation is between the states $\ket{\Phi+}$ and $\ket{\Phi-}$,
in the following way:
\begin{equation}
\rho=\frac{(1-f)}{2}\ket{\Phi+}\bra{\Phi+}+\frac{(1+f)}{2}\ket{\Phi-}\bra{\Phi-},
\end{equation}
 and no more between symmetric
and antisymmetric states as in Equation (\ref{dmf2}).

In conclusion it was shown that the measurement apparatus plays a
central role in the entanglement of identical particles. That
entanglement increases because of broadening in detection
can sound weird at first. However it is important to say
that the detections discussed here are done in a coherent way (as
in most of real cases). If we have modeled the imperfect
detectors as a cluster of perfect detectors things
would be different. Furthermore the relation between identical
particle entanglement (\ie the correlations due exclusively to
indistinguishability) and the entanglement generated by direct
interaction among the particles is not completely understood.
Despite of that the framework presented here can be, \emph{a
priori}, also applied for interacting particle systems and then
could make this relation less obscure.

\acknowledgments{D.C. thanks A.N. de Oliveira and  W. A. T. Nogueira
for useful discussions on Hong-Ou-Mandel interferometry. D.C. and
M.F.S. also acknowledge the Brazilian agency CNPq for financial
support. V.V. acknowledges support from Engineering and Physical
Sciences Council, British Council in Austria and European Union.}

\begin{appendix}
\section{Second order correlation functions as quantum states}
In this appendix we discuss why we consider the second order normalized correlation functions \eqref{dm1}
and \eqref{dmsub} as two-qubit quantum states. Mathematically speaking, any trace $1$ positive operator is
a state on the space state in which it is defined. Physically, however, this seems a poor justification.
Moreover, why one can consider this four-dimensional linear space as a two-qubit system and
obtain conclusions about its entanglement?

To answer this questions we need to consider an interplay of two
aspects of quantum theory: photodetection and post-selection. In a
first quantized notation, if the electromagnetic field state is
described by $\rho$ and we make a one-photon detection described by
the annihilation operator $\op{a}$, the field state after such
measurement will be given by $\rho' = \op{a}\rho\op{a}^{\dagger}$,
properly renormalized (\ie the conditional state). As photodetection
is a destructive measurement, after the click, $\rho'$ describes the
state of the remaining photons. If, however, we have a
non-destructive one-photon detector, this post-selected ``detected
photon'' would be described by the pure Fock state $\ket{1}$ of the
mode annihilated by $\op{a}$. But what if we just have a
non-destructive one-photon detector similar to the one above, except
by the fact that it clicks for any one-photon superposition of two
specific modes (\eg two polarizations of one and the same spatial
mode). This situation is analogous to a degenerate spectrum
projective measurement. The state of the post-selected photon will
preserve the coherence which it possibly had between the two modes,
being described by a linear combination like $\alpha \ket{1,0} +
\beta \ket{0,1}$, or, more generally, by a density operator
$\tilde{\rho} = \op{P}\rho\op{P}$, with $\op{P}$ the projector
$\op{P} = \ket{1,0}\bra{1,0} + \ket{0,1}\bra{1,0}$, and again
properly normalized. Clearly, the state of this two-level system
describes the unresolved mode combination (in the example, the
polarization degree of freedom).

If we detect two spatially distinguishable photons, but each of them with unresolved polarizations,
the first quantized notation becomes cumbersome, but the ideas involved are the same, and the remaining two-qubit
state describes the polarization degree of freedom of this two distinguished photons.

The same reasoning, with two fermions spatially localized ar $r$ and $r'$, but with unresolved spin,
in second quantized notation, give rise to Eq.\eqref{dm1}, that then describes the spin state of
the two localized fermions\cite{Yang}. In this sense, (normalized) second order correlation functions can
be considered as quantum states, and its entanglement properties can be studied, as is done in this paper.
\end{appendix}


\begin{thebibliography}{0}

\bibitem{nielsen}M.A. Nielsen and I.L. Chuang, {\emph{Quantum Computation and
Quantum Information}},(Cambridge Univesity Press, Cambridge, UK,
2000).

\bibitem{vedral}V. Vedral, Cent. Eur. J. Phys. {\bf 2}, 289 (2003).

\bibitem{peres}Asher Peres, \prl{}{\bf{77}}, {1413} {(1996)}.

\bibitem{ohkim}S. Oh, J. Kim, \pra {} {\bf 69}, {054305} {(2004)}.

\bibitem{robust}Guifre Vidal and Rolf Tarrach, \pra {\bf 59},
{141-155}(1999).

\bibitem{negativity}G. Vidal and R.F. Werner, \pra {}{\bf 65},
{032314}{(2002)}.

\bibitem{monot}G. Vidal, \jmo{} {47}, {355} {(2000)}.

\bibitem{horod}M. Horodecki, P. Horodecki and R.
Horodecki, Phys. Lett. A {\bf{223}}, 1-8 (1996).

\bibitem{hongoumandel} C.K. Hong, Z.Y. Ou, and L. Mandel, \prl {\bf{59}} {2044} {(1987)}. S.P. Walborn, A.N. de Oliveira, S. P\'adua and
C.H. Monken, \prl{} {\bf{90}}, {143601}{(2003)}.

\bibitem{Yang}C.N. Yang, \rmp{}{\bf 34}, {694} {(1962)}.

\end{thebibliography}
\end{document}